\documentclass{article}
\usepackage{ijcai22}

\usepackage{times}
\usepackage{soul}
\usepackage{url}
\usepackage[hidelinks]{hyperref}
\usepackage[utf8]{inputenc}
\usepackage[small]{caption}
\usepackage{graphicx}
\usepackage{amsmath, amssymb, amsthm, mathtools, physics, bm, optidef}
\usepackage{float, subcaption,}
\usepackage{csquotes}
\usepackage{multirow, booktabs}
\usepackage{algorithm, algorithmic}
\usepackage{cleveref}
\usepackage{todonotes}
\usepackage{lipsum}
\usepackage{microtype}
\usepackage{acro}
\DeclareAcronym{RL}{short=RL,									short-indefinite=an,		long=reinforcement learning,											long-indefinite=a}
\DeclareAcronym{MARL}{short=MARL,								short-indefinite=a,			long=multi-agent reinforcement learning,								long-indefinite=a}
\DeclareAcronym{MDP}{short=MDP,			short-plural=s,			short-indefinite=an,		long=Markov decision process,						long-plural=es,		long-indefinite=a}
\DeclareAcronym{POMDP}{short=POMDP,		short-plural=s,			short-indefinite=a,			long=partially observable Markov decision process,	long-plural=es,		long-indefinite=a}
\DeclareAcronym{Dec-POMDP}{short=Dec-POMDP,		short-plural=s,			short-indefinite=a,			long=decentralised partially observable Markov decision process,	long-plural=es,		long-indefinite=a}
\DeclareAcronym{POSG}{short=POSG,		short-plural=s,			short-indefinite=a,			long=partially observable stochastic game,	long-plural=s,		long-indefinite=a}
\DeclareAcronym{MAS}{short=MAS,			short-plural=s,			short-indefinite=an,	long=multi-agent system,								long-plural=s,		long-indefinite=a}
\DeclareAcronym{SAS}{short=SAS,			short-plural=s,			short-indefinite=an,	long=single-agent system,								long-plural=s,		long-indefinite=a}
\DeclareAcronym{PMF}{short=PMF,			short-plural=s,			short-indefinite=a,		long=probability mass function,							long-plural=s,		long-indefinite=a}
\DeclareAcronym{PDF}{short=PDF,			short-plural=s,			short-indefinite=a,		long=probability density function,						long-plural=s,		long-indefinite=a}
\DeclareAcronym{CDF}{short=CDF,			short-plural=s,			short-indefinite=a,		long=cumulative density function,						long-plural=s,		long-indefinite=a}
\DeclareAcronym{AI}{short=AI,									short-indefinite=an,	long=artificial intelligence,							long-plural=s,		long-indefinite=an}
\DeclareAcronym{DFA}{short=DFA,			short-plural=s,			short-indefinite=a,		long=discrete finite automaton,							long-plural-form=discrete finite automata,		long-indefinite=a}
\DeclareAcronym{MLE}{short=MLE,			short-plural=s,			short-indefinite=an,	long=maximum likelihood esitmation,						long-plural=s,		long-indefinite=a}
\DeclareAcronym{MAP}{short=MAP,			short-plural=s,			short-indefinite=a,		long=maximum a posteriori,								long-plural=s,		long-indefinite=a}
\DeclareAcronym{IID}{short=IID,			short-plural=s,			short-indefinite=an,	long=identically and independently distributed,			long-plural=s,		long-indefinite=an}
\DeclareAcronym{MCTS}{short=MCTS,		short-plural=s,			short-indefinite=an,	long=Monte-Carlo tree search,							long-plural=s,		long-indefinite=a}
\DeclareAcronym{FST}{short=FST,									short-indefinite=a,		long=few-shot teamwork,													long-indefinite=a}
\DeclareAcronym{AHT}{short=AHT,									short-indefinite=an,	long=ad hoc teamwork,													long-indefinite=a}

\DeclarePairedDelimiterX{\infdivx}[2]{(}{)}{#1\;\delimsize\|\;#2}

\newcommand{\StateSpace}{\mathcal{S}}

\newcommand{\ObsFunc}{O}
\newcommand{\ObsSpace}{\Omega}

\newcommand{\ActionSpace}{\mathcal{A}}
\newcommand{\StateTransFunc}{T}

\newcommand{\RewardFunc}{R}

\newcommand{\Policy}{\pi}

\newcommand{\Task}{\mathcal{T}}
\newcommand{\nSourceTask}{M}
\newcommand{\iSourceTask}{m}

\newcommand{\iTarget}{T}

\newcommand{\Team}{P}
\newcommand{\LearnAlg}{\mathbb{L}}
\newcommand{\SourceStage}{S}
\newcommand{\AdjStage}{A}
\newcommand{\BaselineTarget}{\odot}

\title{Few Shot Teamwork}

\author{
Elliot Fosong
\and
Arrasy Rahman\and
Ignacio Carlucho\And
Stefano V. Albrecht
\affiliations
School of Informatics, University of Edinburgh\\
\emails
\{e.fosong, arrasy.rahman, ignacio.carlucho, s.albrecht\}@ed.ac.uk
}

\begin{document}

\maketitle

\begin{abstract}
	We propose the novel \emph{\acf{FST}} problem, where skilled agents trained in a team to complete one task are combined with skilled agents from different tasks, and together must learn to adapt to an unseen but related task.
	We discuss how the \ac{FST} problem can be seen as addressing two separate problems:
	one of reducing the experience required to train a team of agents to complete a complex task;
	and one of collaborating with unfamiliar teammates to complete a new task.
	Progress towards solving \ac{FST} could lead to progress in both \acl*{MARL} and \acl*{AHT}.
\end{abstract}

\section{Introduction}
In this paper, we introduce the \emph{\acf{FST}} problem, wherein subteams of agents skilled with respect to different tasks are combined into a larger team, and must learn to adapt to an unseen but related task.
An illustrative example of the \ac{FST} problem is a football (soccer) setting, in which two skilled defenders are trained in defensive drills, and are combined with two skilled attackers, with whom they must learn to coordinate to play 5-a-side football.
We motivate this problem by considering two related problems, which can be modelled by the \ac{FST} framework.
In the first problem, \ac{FST} can be viewed through a \emph{curriculum learning} framing, where the goal is to accelerate training a team to complete a complex task.
In the second problem, \ac{FST} can be viewed through an \emph{\acl{AHT}} framing, where agents collaborate with previously unencountered teammates to complete a new task.

\textbf{Training a team for a complex task.} (Curriculum Framing).
Training a team of agents to complete a complex task can be difficult.
Challenges include
policy search in a large joint policy space, 
the multi-agent credit assignment problem,
and the non-stationarity presented by mutually adapting agents.
A possible solution to these problems is to design a curriculum for the team by decomposing the complex target task into simpler subtasks.
Subsets of the team could be trained on the subtasks, where they can gain skills relevant to the target task.
These skilled subteams are combined and fine-tuned on the complex target task,
hopefully requiring significantly less training than learning from scratch, as agents need not relearn the capabilities developed in the simpler subtasks.
In the 5-a-side football example, we aim to accelerate learning the complex 5-a-side task by first training the defenders and attackers in simpler defensive and offensive drills respectively, then combining them to form and train a full 5-a-side team.

\textbf{Collaborating with unfamiliar agents on a new task} (\ac{AHT} framing).
Having agents with existing skills learn to quickly collaborate with previously unencountered teammates to complete a new task is an unsolved problem.
Agents do not know a priori how to coordinate with the new teammates,
must learn to cooperate despite non-stationarity,
and do not know how to complete the new task, even though task completion benefits from agents' existing skills.
Returning to the 5-a-side football example, we could consider the defenders/attackers as being selected from separate training academies, and must learn to play with their new teammates.
Design knowledge transfer mechanisms
could promote coordination by enabling agents to communicate their capabilities and intended behaviours, improving the legibility of their actions to other agents;
or helping settle disputes about which joint action should be taken;
or transferring useful skills between agents.

By solving the \ac{FST} problem, we could expect several benefits related to \ac{MARL} and \acf{AHT}.
Solutions could lead to greatly reduced computational or experience requirements for training a multi-agent team to solve complex but decomposable tasks.
Progress in \ac{FST} could also represent progress towards agents capable of long-term autonomy, which are able to work together with other agents to complete a novel task.

\section{Problem Definition}\label{sec:ref}

	The \ac{FST} problem consists of two stages:
	a \emph{source stage}, where teams of agents are trained with respect to their initial, simpler source tasks,
	and an \emph{adjustment stage}, where teams from the source stage are combined and fine-tuned to complete the more complex target task.
	We model a set of $\nSourceTask$ source tasks, $\Task_1, \dots, \Task_\nSourceTask$,
	and a single target task $\Task_\iTarget$.
	Each task $\Task$ is a common-payoff task, and can be modelled by a \acs{Dec-POMDP}:
	\begin{equation}
		\Task = \langle \Team, \StateSpace, \qty{\ActionSpace_i}_{i\in\Team}, \StateTransFunc, \RewardFunc, \qty{\ObsSpace_i}_{i\in\Team}, \ObsFunc, \gamma\rangle,
	\end{equation}
	Where
	$\Team$ is the set of agents (team);
	$\StateSpace$ is the state space;
	$\ActionSpace_i$ is the action space for agent $i$;
	$\StateTransFunc$ is the state transition probability density function;
	$\RewardFunc$ is the reward function;
	$\ObsSpace_i$ is the observation space for agent $i$;
	$\ObsFunc$ is the observation probability density function;
	and
	$\gamma$ is the discount factor.

	\subsection{Source Stage}
	In the source stage, each subteam is trained with respect to a source task, where they learn skills relevant to the more complex target task.
	For each of the $\nSourceTask$ source tasks, $\Task_\iSourceTask$, the team $\Team_\iSourceTask$ trains for $N_\SourceStage$ training steps using learning algorithm $\LearnAlg_\SourceStage$, generating a joint task policy $\Policy_\iSourceTask$,
	and optionally other data or functions derived from the training process, such as a buffer of source stage experiences, which might be useful during the adjustment stage.

	\subsection{Adjustment Stage}
	In the adjustment stage, subteams are combined to form the target task's team,
	$\Team_\iTarget \subseteq \bigcup^\nSourceTask_{\iSourceTask}{\Team_\iSourceTask}$,
	and then learn to coordinate by practising in the target task for a limited number of training steps, $N_\AdjStage$,
	after which the team is evaluated.
	During this stage, agents use learning algorithm, $\LearnAlg_\AdjStage$,
	designed to promote coordination and exploration of the new task
	using the skilled source stage policies $\Policy_\iSourceTask$ and other data or functions derived during the source stage.
	At the end of the $N_\AdjStage$ training steps, the performance of the new team is evaluated on task $\Task_\iTarget$, which forms the optimisation objective of the \ac{FST} problem.

	\subsection{Objective}\label{sec:obj}
	The overall objective of the \ac{FST} problem is to maximise the mean returns of the final team on the target task $\Task_\iTarget$ after the $N_\AdjStage$ adjustment stage training steps.
	The main parameters to optimise are
	the source and adjustment stage learning algorithms, $\LearnAlg_\SourceStage$ and $\LearnAlg_\AdjStage$ respectively.
	Which of the curriculum or \acl{AHT} framings we adopt determines our settings for the number of training steps, $N_\SourceStage$ and $N_\AdjStage$.\\
	In the curriculum framing, we simply desire to find solutions which require less training experience to achieve desired performance on the target task than \ac{MARL} baselines.
	If \ac{MARL} baselines require $N_\BaselineTarget$ training steps to reach performance $G_\BaselineTarget$ from scratch on task $\Task_\iTarget$,
	then a successful approach to \ac{FST} should reach performance $G_\BaselineTarget$ with $\nSourceTask \times N_\SourceStage + N_\AdjStage\ll N_\BaselineTarget$.\\
	In the \ac{AHT} framing, we are less concerned with $N_\SourceStage$, which may be large, but wish to minimise the $N_\AdjStage$ required to reach target performance $G_\BaselineTarget$.
	We may aim for $N_\AdjStage$ to be some small absolute number (e.g., in the tens or hundreds),
	whereas under the curriculum framing, for sufficiently complex tasks we may tolerate large values of $N_\AdjStage$ (e.g., upwards of millions) provided this is relatively small compared to $N_\BaselineTarget.$

	\subsection{Assumptions}\label{sec:ass}
	There are two key assumptions we make to constrain the \ac{FST} problem.
	Firstly, we assume that during the source stage, teams are unaware of other source tasks and the target task,
	so information from other tasks cannot be used by the learning algorithm $\LearnAlg_\SourceStage$.
	This assumption is justified in the \ac{AHT} framing, but could be relaxed in the curriculum framing.
	We choose to make this assumption to simultaneously target both framings. \\
	Secondly, we assume a relationship between the source and target tasks, such that some skills which are useful in source tasks are useful in the target task.
	We attempt to formalise this relationship by stating that for some subset of the state space, the optimal joint policy for the target task has some agents acting according to the optimal policy of the source task.
	Such assumptions are necessary --- No Free Lunch theorems suggest that without regularity assumptions, generalisation is not possible \cite{wolpertNoFreeLunch1997}.
	Alternative forms of the source--target relationship assumption could be explored in future work.

\section{Differences Between the Two \acs{FST} Framings}
	The main difference between the curriculum and \ac{AHT} framings is that there is a different emphasis in the objective function, discussed in \cref{sec:obj}.
	Rather than a categorical distinction between the two framings, \ac{FST} encompasses a spectrum of problems with strict versions of the framings at each end.
	There are, however, differences between assumptions made by the two approaches.
	As discussed in \cref{sec:ass}, the curriculum framing need not assume lack of access to the target task during source training.
	The \ac{AHT} framing could assume that the source tasks are given, whereas in the curriculum framing, the source tasks could be a parameter to choose.
	Initially, we intend to assume a method exists for choosing source tasks, and hand pick source tasks during experiments.
	Finally, despite their similarity, the two framings may require very different solutions.
	Approaches to the \ac{AHT} framing will need to employ stronger inductive biases than curriculum approaches, due to the more extreme few-shot nature of the objective.

\section{Related Work}
	\Acf{AHT} is the problem of coordinating on the fly with previously unseen teammates \cite{mirskySurveyAdHoc2022,stoneAdHocAutonomous2010}.
	Similarly to \ac{FST}, in \ac{AHT}, agents need to adapt to other agents whose behaviours are initially unknown.
	However, unlike \ac{FST}, \ac{AHT} does not consider task transfer.
	Due to the added complication of task transfer, we allow for slower adaptation in \ac{FST} solutions than we would normally accept in \ac{AHT}.
	Common \ac{AHT} approaches such as type-based reasoning \cite{albrechtReasoningHypotheticalAgent2017,albrechtBeliefTruthHypothesised2016,barrettTeamworkLimitedKnowledge2013} are unlikely to apply to \ac{FST},
	as they typically assume teammates will be drawn from a known type distribution, whereas in \ac{FST} we assume agents have no knowledge of the other source tasks.
	Solving the \ac{FST} problem could extend progress in \ac{AHT} to consider cases where there is also task transfer, and possibly lead to the development of alternatives to the dominant type-based reasoning approaches.

	A related research area is transfer learning \cite{silvaSurveyTransferLearning2019}. 
	There are two types of transfer learning relevant to \ac{FST}: knowledge transfer between agents in a single task; and transferring existing skills between different tasks.
	The former problem shares the same \enquote{distribution of skills} property as \ac{FST} has a range of approaches, including \emph{learning to teach} \cite{omidshafieiLearningTeachCooperative2019}.
	However, these approaches are not designed to consider transfer between tasks.

	Curriculum learning is another related field, which trains agents to complete complex tasks by providing them with a series of progressively more challenging tasks \cite{narvekarCurriculumLearningReinforcement2020a}. 
	This is similar to the curriculum framing of \ac{FST}.
	However, much of this work considers single-agent \acs{RL}, and focusses on choosing source tasks for the curriculum \cite{dennisEmergentComplexityZeroshot2020}, which we postpone to future work.

\bibliographystyle{named}
\bibliography{fst}

\end{document}